\documentclass[aps, amsmath, amssymb, reprint, floatfix, superscriptaddress]{revtex4-2}

\usepackage{graphicx}
\usepackage{dcolumn}
\usepackage{bm}
\usepackage{xcolor}

\begin{document}


\title{\mbox{AlGaN}/AlN heterostructures: an emerging platform for nonlinear integrated photonics}

\author{Sinan Gündo\u{g}du}
 \email{sinang@physik.hu-berlin.de}
 \affiliation{Department of Physics, Humboldt-Universit\"{a}t zu Berlin, Newtonstr. 15, 12489 Berlin, Germany}%
 \affiliation{Ferdinand-Braun-Institut (FBH), Gustav-Kirchhoff-Str. 4, 12489 Berlin, Germany}

\author{Sofia Pazzagli}%
\affiliation{Department of Physics, Humboldt-Universit\"{a}t zu Berlin, Newtonstr. 15, 12489 Berlin, Germany}%

\author{Tommaso Pregnolato}
\affiliation{Department of Physics, Humboldt-Universit\"{a}t zu Berlin, Newtonstr. 15, 12489 Berlin, Germany}%
 \affiliation{Ferdinand-Braun-Institut (FBH), Gustav-Kirchhoff-Str. 4, 12489 Berlin, Germany}

\author{Tim Kolbe}
 \affiliation{Ferdinand-Braun-Institut (FBH), Gustav-Kirchhoff-Str. 4, 12489 Berlin, Germany}
 
\author{Sylvia Hagedorn}
 \affiliation{Ferdinand-Braun-Institut (FBH), Gustav-Kirchhoff-Str. 4, 12489 Berlin, Germany}
 
\author{Markus Weyers}
 \affiliation{Ferdinand-Braun-Institut (FBH), Gustav-Kirchhoff-Str. 4, 12489 Berlin, Germany}

\author{Tim Schr\"{o}der}
\affiliation{%
Department of Physics, Humboldt-Universit\"{a}t zu Berlin, Newtonstr. 15, 12489 Berlin, Germany}%
 \affiliation{Ferdinand-Braun-Institut (FBH), Gustav-Kirchhoff-Str. 4, 12489 Berlin, Germany}

\date{\today}

\begin{abstract}
In the rapidly evolving area of integrated photonics, there is a growing need for materials that satisfy the particular requirements of increasingly complex and specialized devices and applications. Present photonic material platforms have made significant progress over the past years; however, each platform still faces specific material and performance challenges. We introduce a novel material for integrated photonics: Aluminum Gallium Nitride (\mbox{AlGaN}) on Aluminum Nitride (AlN) as a platform for developing reconfigurable and nonlinear on-chip optical systems. \mbox{AlGaN} combines compatibility with standard semiconductor fabrication technologies,  high electro-optic modulation capabilities, and large nonlinear coefficients while providing a broad and low-loss spectral transmission range, making it a viable material for advanced photonic applications. In this work, we design and grow \mbox{AlGaN}/AlN heterostructures and integrate fundamental photonic building blocks into these chips. In particular, we fabricate edge couplers, low-loss waveguides, directional couplers, and tunable high-quality factor ring resonators to enable nonlinear light-matter interaction and quantum functionality. The comprehensive platform we present in this work paves the way for nonlinear photon-pair generation applications, on-chip nonlinear quantum frequency conversion, and fast electro-optic modulation for switching and routing classical and quantum light fields.
\end{abstract}

\maketitle

\section{Introduction}

\begin{figure*}
    \centering
    \includegraphics[width=\textwidth]{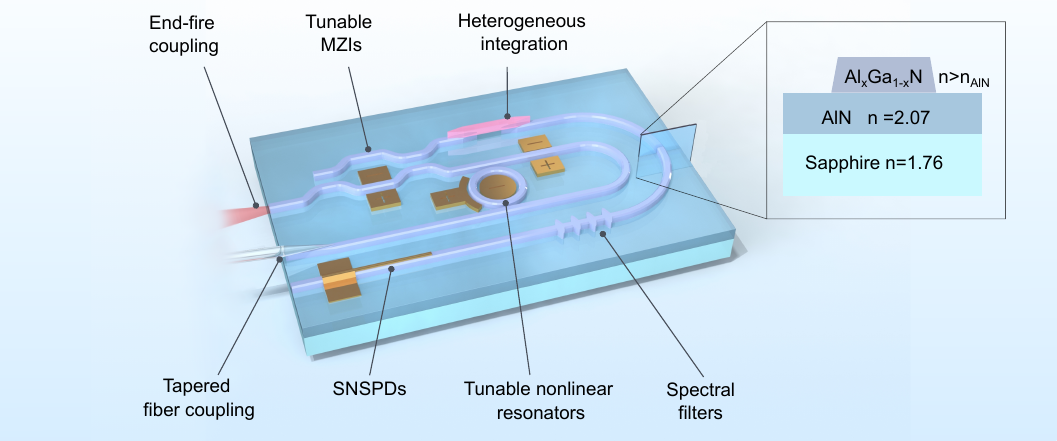}
    \caption{\textbf{Illustration of \mbox{AlGaN}/AlN platform for photonic integrated circuits.} The \mbox{AlGaN}/AlN platform consists of various passive and active components. The core innovation is the photonic layer in which light is guided by engineering a few hundred-nanometer-thick high refractive index top layers, separating the optical mode from the sapphire substrate interface. Through lateral patterning, a large variety of passive and active optical elements can be integrated, including waveguides, directional couplers, tunable Mach-Zehnder interferometers, microring resonators as spectral filters and nonlinear devices, photonic crystals, and superconducting nanowire single-photon detectors. Additionally, this platform allows for the heterogeneous integration of functional electronic and optically active components, such as laser diodes, and solid-state quantum light sources, such as defects in diamonds and fluorescent molecules in organic matrices. }
    \label{fig:ppconcept}
\end{figure*}

Advanced classical and quantum photonic applications, such as photonic neuromorphic computing \cite{Shastri2021}, quantum sensing \cite{Xavier2021} and quantum networking \cite{Knaut2023, Zhang2019,Onewayrepeater,Multipartite,Coherentcontrol} rely on photonic integrated circuits that enable the compact, efficient, and high-rate implementation of a variety of optical functionalities. In addition to functionalities adapted from stand-alone optical devices, integrated circuits provide access to mode-multiplexing or routing \cite{Liu2019,9552852} and efficient fiber coupling. A comprehensive photonic device platform  \cite{Siew2021}, therefore, integrates various components, such as low-loss waveguides, efficient directional couplers, spectral filters, and tunable Mach-Zehnder interferometers (Fig.~\ref{fig:ppconcept}). A tunable ring resonator is another particularly versatile component, which can be employed to spectrally filter different modes, to enhance light-matter interaction, or to enable nonlinear frequency conversion and photon-pair generation.\cite{Logan2018,Guo2016,Wang2023}. Additionally, given the large variety of electronic and optical components readily available on different material platforms, like detectors \cite{SNSPD_AlN, EsmaeilZadeh2021} or quantum light sources \cite{Mouradian2015,Pregnolato2024}, heterogenous integration is a key approach for building photonic integrated chips with a high level of complexity. An ideal photonic platform should, therefore, facilitate such a method while also exhibiting properties such as low optical losses, fast electro-optic modulation, and significant optical nonlinearities to route and manipulate the propagating photons with high efficiency.

Different material platforms are currently being studied, each with its own advantages and disadvantages in meeting the requirements of a versatile, scalable, and reliable photonic platform. For the ultraviolet and visible spectral range, there are several established alternatives to silicon, which is not transparent in this range.  Lithium niobate on insulator (LNOI) is one of the leading platforms due to its outstanding electro-optic coefficient, high optical nonlinearity, and recent progress in technological development \cite{LNBOI-Poberaj2012,Qi2020,Saravi2021,Zhu2021}. However, LNOI faces several challenges. Non-standard etching processes are needed to fabricate optical components. These processes are incompatible with the conventional manufacturing standards, potentially hindering its scalability \cite{LNB-Han2023}. Moreover, despite the advances in epitaxial techniques, research, and industry still predominantly rely on smart-cut lithium niobate derived from bulk crystals, which typically suffer from crystal damage. Furthermore, the limited thickness range ($>$300 nm) of smart-cut lithium niobate restricts some applications, such as nanoscale integrated optics \cite{Zivasatienraj2021}. 

Silicon nitride (SiN), another established material platform, offers low losses \cite{Buzaverov2023} and is compatible with silicon manufacturing standards. SiN-based active optical elements, however, such as tunable Mach-Zehnder Interferometers (MZIs), typically rely on thermo-optic modulation due to their limited electro-optic functionality \cite{Hermans:19}. Thermo-optic control is comparably slow and generally limited to kHz modulation rates \cite{Nejadriahi:20}. Another nitride-based material, aluminum nitride (AlN) on sapphire, has been explored due to its substantial nonlinear coefficient, its broad transparency range, and its moderate electro-optic coefficient \cite{Li2021, Sun2019, Wu2020, Liu2018, Iu2017}. AlN on sapphire, however, has yet other limitations. It is usually grown by sputtering or metal-organic vapor phase epitaxy
(MOVPE), typically on a sapphire substrate. Consequently, the AlN layer suffers from a high defect density at the sapphire interface \cite{Hagedorn2019, HRTEM_TOKUMOTO20094886, Prism_coupling_Dogheche2000}, which leads to relatively high optical scattering and absorption losses. Furthermore, commercially available AlN suffers from non-negligible thickness variation, making it challenging to fabricate photonic components scalably. 

In this study, we introduce a platform for photonic integrated circuits based on aluminum gallium nitride (\mbox{AlGaN}), where the \mbox{AlGaN} layer acts as a guiding photonic layer that can be patterned into a large variety of optical components. This approach overcomes challenges in AlN on sapphire platforms, such as crystal defects at the sapphire-AlN interface and lateral thickness variation, and introduces further material-specific functionality. With the additional \mbox{AlGaN} layer, we reduce optical losses and improve device performance compared to AlN on sapphire by shifting the optical mode away from this interface. The ability to control the ratio of aluminum to gallium in the ternary alloy Al$_x$Ga$_{1-x}$N allows for tailored properties of the photonic layer, such as bandgap and refractive index engineering. Moreover, \mbox{AlGaN}'s wide band gap enables efficient light transmission across a broad spectrum, depending on composition from below 250 nm to long-wave infrared. Doping \mbox{AlGaN} for electrical conductivity combines optical and electronic functionalities, enabling, for example, current-induced refractive index change \cite{Schwarz2004, Bulutay2010}. 

Our \mbox{AlGaN} platform benefits from the advanced state-of-the-art of GaN, which has, over the past decade, evolved into a foundational material for electronic and electro-optic devices, including power electronics and light generation \cite{GaNPower10128694,GanLedATabbakh2023}. For example, GaN LEDs and lasers are widely used for producing blue and green light \cite{BlueLiang2021,GreenHu2020}. The epitaxial growth of GaN/\mbox{AlGaN}/InGaN heterostructures on sapphire templates, a well-established practice, opens up possibilities for integrating light sources and electronics on the photonic platform.

\begin{figure*}[t!]
    \centering
    \includegraphics[width=\textwidth]{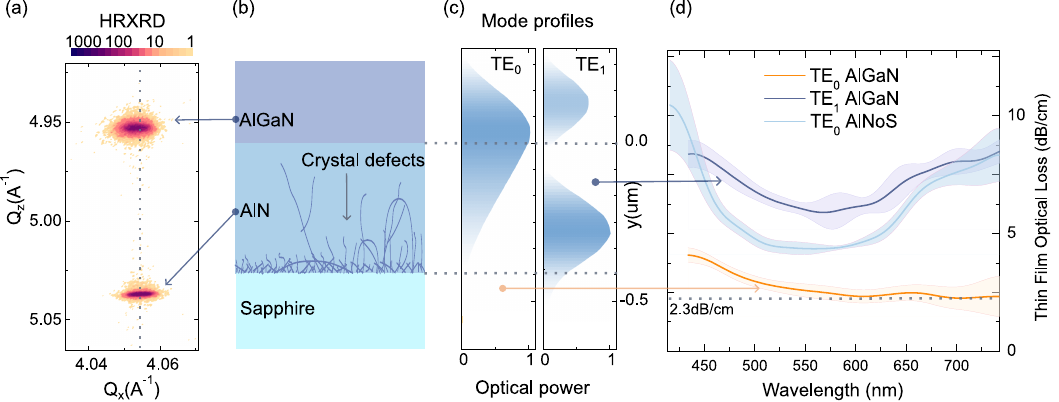}
    \caption{\textbf{Material characterization.} (a) High-resolution X-ray diffraction reciprocal space maps (11-24 reflection) illustrate the strain relaxation in the \mbox{AlGaN} and AlN layers. Two distinct peaks correspond to the AlN and \mbox{AlGaN} layers, respectively. Analysis of these peaks reveals an Al-mole fraction x = 0.69 and a mere 3$\%$ relaxation of in-plane compressive strain in the \mbox{AlGaN} layer, indicating a predominantly pseudomorphic growth on the AlN template and avoiding the generation of strain-relief defects. (b) Illustration of the basic heterostructure on a sapphire substrate. (c) Supported  TE$_0$ and TE$_1$ modes for a 250 nm thick \mbox{AlGaN} layer on AlN at 632 nm. (d) The optical loss spectrum of these modes in the \mbox{AlGaN}/AlN heterostructure and the TE$_0$ mode of a 0.4 µm thick AlN single layer on sapphire film (AlNoS), measured by prism coupling, where shaded areas indicate the uncertainty bounds. 
}
    \label{fig:algan}
\end{figure*}

The inherent properties of \mbox{AlGaN}, such as its relatively high electro-optic coefficient \cite{AlGaN_suscept_meas, Chen1995}, make it an ideal candidate for rapid modulation and the development of reconfigurable optical devices. Its significant second-order optical nonlinearity paves the way for on-chip nonlinear photonic devices, including parametric oscillators and sum or difference frequency generators. The generation of correlated photon pairs in the ultraviolet (UV) and visible spectrum has been proposed using an \mbox{AlGaN}/AlN integrated photonics platform via spontaneous four-wave mixing (SFWM) in an \mbox{AlGaN} microring resonator \cite{AlGaNRing}. \mbox{AlGaN} has also been explored as a potential platform for stimulated Brillouin scattering devices such as a racetrack Brillouin laser \cite{Brillouin}.

To the best of our knowledge, only very few studies have used \mbox{AlGaN} as a photonic material. Li et al. \cite{FabPaperLi2004} fabricated large, multimode waveguides with an AlN/GaN multiple quantum well core and Al$_{0.1}$Ga$_{0.9}$N claddings. Although this is a notable achievement, the large size and relatively high sidewall roughness do not enable application for single-mode photonics. In addition, the wavelength range is limited to around 350 nm and longer. Bruch et al. \cite{FabPaperBruch2017} fabricated ring resonators on Al$_{0.1}$Ga$_{0.9}$N nanomembranes transferred onto SiO$_2$/Si substrates. While their work demonstrated the versatility of \mbox{AlGaN}, using nanomembranes introduces difficulties with handling and scalability. Recently, Shin et al. showcased the enhancement of the Pockels effect in multiple \mbox{AlGaN}/AlN quantum wells on AlN due to the large built-in polarization of the quantum wells \cite{EnhancedPockels}. Despite encountering relatively high propagation losses (18.3 dB/cm), their work achieved a 20-fold increase in second-order susceptibility compared to bare AlN. This work shows the capability of the \mbox{AlGaN}/AlN as a platform for cutting-edge engineering to achieve nonlinearities beyond those available in bulk materials.

Here, we develop heterogeneously grown \mbox{AlGaN} into a photonic platform. To fabricate photonic devices, we grow Al$_x$Ga$_{1-x}$N, with x$=0.69$, directly on high-temperature annealed AlN-on-sapphire templates with a reduced dislocation density \cite{Hagedorn2020}. Into this heterostructure, we subsequently fabricate the fundamental constituents of photonic integrated circuits with electron beam lithography and plasma etching: waveguides, directional couplers, ring resonators, and fiber couplers. Finally, we present the design of a device to achieve efficient and flexible phase matching for entangled photon generation using Spontaneous Parametric Down-Conversion (SPDC) at telecom wavelengths employing such a heterostructure. 

\section{Results}
\subsection{Wafer Growth and Characterization}

The core ingredient of our novel integrated photonics platform is an \mbox{AlGaN} heterostructure (Fig. \ref{fig:algan}(b)). The AlN/\mbox{AlGaN} layer stack is epitaxially grown on a c-plane-oriented sapphire substrate with an offcut of 0.25° towards an m-plane. First, a 350 nm thick AlN layer is deposited using epitaxial magnetron sputtering. The sputtered material's threading dislocation density (TDD) was decreased through high-temperature annealing (HTA), following the process described by Miyake et al. \cite{Miyake2016}. During HTA, a temperature of 1700 °C is maintained for a relatively short duration of 1h to prevent the formation of aluminum oxynitride on the AlN surface \cite{Hagedorn2019}. Subsequently, the rough sputtered and annealed surface is smoothened by MOVPE growth of 50 nm AlN in a step flow growth regime. The AlN layer, with a total thickness of 400 nm, exhibits a TDD of \(7.5 \times 10^8 \, \text{cm}^{-2}\), as estimated from high-resolution X-ray diffraction (HRXRD) measurements of the symmetric 0002 and skew-symmetric 10-12 \(\omega\)-rocking-curves \cite{Pantha2007}. Atomic force microscopy (AFM) revealed a surface RMS roughness of 0.09 nm over a \(25 \, \mu\text{m}^2\) area. 

Next, a 250 nm thick \mbox{AlGaN} waveguide layer is grown by MOVPE in a 6 x 2-inch close-coupled showerhead reactor on the AlN template, utilizing trimethylaluminum, triethylgallium, and ammonia as source materials, with hydrogen as the carrier gas, and a growth temperature of 1015 °C. In-situ reflectometry measurements verify the attainment of the targeted layer thickness of 250 nm. After \mbox{AlGaN} growth, we measure the surface RMS roughness by AFM to be 0.5 nm. This smooth surface supports integrating photonic devices as introduced here and enables the hetero-integration of superconducting nanowire single-photon detectors (SNSPDs) \cite{SNSPD_AlN} and quantum devices.
The strain relaxation towards the underlying AlN buffer and the Al-mole fraction x of the \( \text{Al}_{x}\text{Ga}_{1-x}\text{N} \) layers is determined by HRXRD using $\omega -\omega/2\Theta$  reciprocal space maps (RSM) of the 11-24 reflection. Each RSM shows two sharp peaks (Fig. \ref{fig:algan}(a)), corresponding to the AlN template and \mbox{AlGaN} layers. By analyzing these data, an Al-mole fraction x of 0.69 and a relaxation of the in-plane compressive strain of only 3$\%$ can be determined for the \mbox{AlGaN} layer. This shows that the \mbox{AlGaN} grew almost pseudomorphically on the AlN template, and the generation of misfit dislocations for strain relief can thus be largely excluded.

\subsection{Optical Material Properties}
We use the prism coupling technique as described in \cite{Prism1973Weber:73,PrismAlNDogheche1999} to quantify the optical losses in the as-grown thin films. It utilizes a rutile prism to couple a tunable white laser (SuperK Fianium) into the heterostructures. The losses are quantified by monitoring the decay of the scattered light with a digital camera. We characterize and compare a 400 nm AlNoS sample and an \mbox{AlGaN}/AlN/sapphire wafer. The AlNoS supports a single TE$_0$ mode at 630 nm, whereas the \mbox{AlGaN} structure exhibits two distinct modes, TE$_0$ and TE$_1$, at the same wavelength. By fine-tuning the coupling angle, we isolate and analyze the losses of these modes, as reported in Fig. \ref{fig:algan}(d).\\ 
The TE$_0$ mode, in the \mbox{AlGaN} structure, exhibits optical losses that decrease with increasing wavelength, reaching a minimum value of 2.3 dB/cm above 600 nm. In contrast, the TE$_1$ mode of the \mbox{AlGaN} and the TE$_0$ mode of the AlNoS exhibited higher losses, with minimum values of approximately 5.9 and 4.3 dB/cm, respectively, which increase again for wavelengths over 625 nm. It is worth noting that both of these modes are localized at the interface between two different materials and, thus, are sensitive to the interface's inherent properties. Besides the roughness at the interface, additional sources of loss could be due to defects and color centers in AlN, which are presumably concentrated at the AlN-sapphire interface as shown by TEM measurements by \cite{Hagedorn2019,HRTEM_TOKUMOTO20094886,Prism_coupling_Dogheche2000}. 
AlN has defects comprising a mix of impurities, vacancies, and lattice defects. Factors like threading dislocations influence the formation of these defects, although the exact conditions for their formation are yet not fully understood \cite{AlNColor_Cannon2023, Varley2016,Xue2022}.

\begin{figure}
    \centering
    \includegraphics[width=0.45\textwidth]{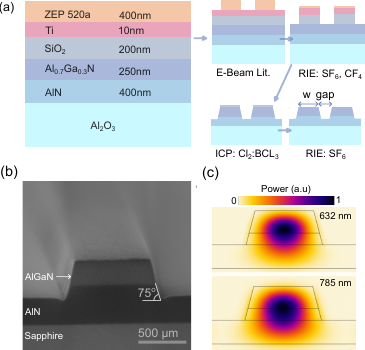}
    \caption{\textbf{Fabrication geometries.} (a) Epitaxial and mask layers for the fabrication of the optical nanowaveguides. (b) SEM image of a cleaved facet of a waveguide, where the different materials can be clearly identified. (c) Simulated optical power distribution of the fundamental TE mode of the waveguides at 632 nm and 785 nm.}
    \label{fig:devices}
\end{figure}

\subsection{Device Fabrication}

In the fabrication process of waveguides and ring resonators, we initially deposit a 200 nm thick SiO$_2$ layer using Plasma Enhanced Chemical Vapor Deposition (PECVD) to serve as a hard mask. We then coat a 10 nm thick Ti layer via e-beam evaporation to act as a charge dissipation layer, followed by 400 nm ZEP 520A e-beam lithography resist. The process steps are illustrated in Fig. \ref{fig:devices}(a). After patterning the resist, we transfer the pattern to the hard mask using a two-step reactive ion etching (RIE) process: we utilize SF$_6$ plasma to etch Ti and CF$_4$ plasma to etch SiO$_2$. In the final etching step, we process the \mbox{AlGaN} layer using BCl$_3$:Cl$_2$:He at a ratio of 10:50:10 sccm, under 600 W ICP power and 100 W RF power, at a pressure of 1 Pa. The \mbox{AlGaN}-to-SiO$_2$ etch rate ratio is approximately 3/1. We implement a laser-cut notch at the chip's edge to facilitate precise cleaving. Fig. \ref{fig:devices}(b) presents an SEM image of a cleaved facet, revealing the AlN and \mbox{AlGaN} layers. We measure the waveguides to have a wall angle of 75$^\circ$. Fig. \ref{fig:devices}(c) shows the simulated power profiles of the fundamental TE mode of the waveguides at wavelengths of 632 nm and 785 nm. At the measurement wavelength of 785 nm used for our ring resonators, 90$\%$ of the optical power is located above the AlN-sapphire interface. At a shorter wavelength of 632 nm, this number increases to 95$\%$, indicating that most of the optical power is well separated from the lossy boundary at the sapphire interface.
\begin{figure*}[ht]
    \centering
    \includegraphics[width=\textwidth]{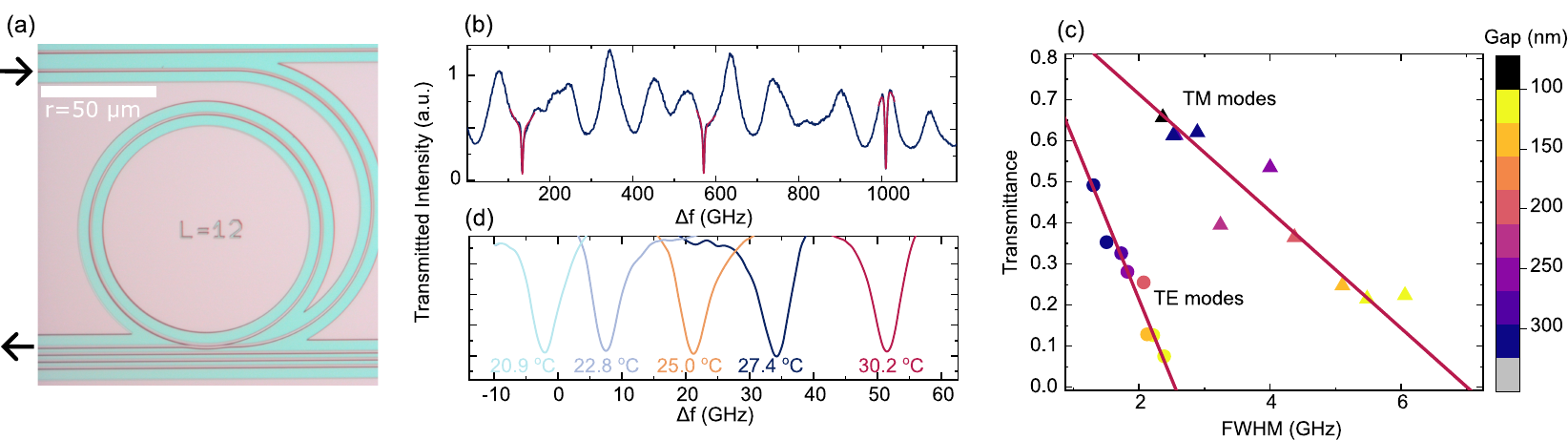}
    \caption{\textbf{ Ring resonator and waveguide device characterization. }(a) Optical microscope image of a waveguide coupled ring resonator. (b) The transmission through a ring resonator, with a width of 580 nm and a 100 nm coupling gap, is plotted as a function of laser detuning. The dark red lines represent fits using a Lorentzian function with a quadratic baseline. (c) The graph of transmittance versus the full width at half maximum (FWHM) of resonances for different coupling gaps between the semi-racetrack waveguide and the ring. The red lines represent linear fits for TE and TM modes. (d) The resonance frequency tuning with temperature.}
    \label{fig:ring}
\end{figure*}

\subsection{Tunable Microring Resonators}

We fabricate microring resonators with a radius of 50 µm and a top width of 580 nm, coupled to racetrack waveguides with the same width and cleaved at both ends, as depicted in  Fig. \ref{fig:ring}(a). The sample is mounted on a 3D piezoelectric nanopositioner (Physik Instrumente, NanoCube) and characterized with a custom-made inverted microscope, where the same objective (Zeiss, LD EC Epiplan-Neofluar, 20X, NA = 0.22) is used to launch the laser light into the input waveguide facet and collect it from the output one. The alignment of the incoupling light is monitored with an EMCCD camera (Andor, iXon Ultra 897) and optimized by maximizing the intensity of the outcoupled signal by adjusting both sample position and laser incoupling angle.  Propagation losses are then inferred by measuring the intrinsic quality factor of the microring resonators. To reach critical coupling,  which allows for the estimation of resonator losses, we fabricate several microring resonators coupled to racetrack waveguides with different coupling gaps, ranging from 100 to 325 nm, as displayed in Fig. \ref{fig:ring}(c). We measure the transmission of a tunable 785 nm laser (Toptica, DFB PRO centered at 784.6 nm, MHF over 2 nm) and use a linear film polarizer (Thorlabs, LPVIS) to excite and collect TE or TM modes selectively. Fig. \ref{fig:ring}(b) illustrates the typical transmission as a function of the laser's frequency, with a baseline primarily attributed to the interference in measurement optics. This modulation can be described with a quadratic baseline from which the resonant peaks can be discriminated and fitted with a Lorentzian function, as indicated in red in Fig. \ref{fig:ring}(b). We extrapolate the resonance widths at the critical coupling, i.e., zero transmittance. The key parameters, such as the critical coupling FWHM, free spectral ranges (FSRs), and intrinsic Q-factors for TE and TM modes, are summarized in Table \ref{tab:parameters}.

\begin{table}[h]
    \centering
    \caption{Summary of the measured and calculated optical and thermo-optic parameters for the Al$_x$Ga$_{1-x}$N (x=0.69) ring resonators with 50 $\mu$m radius.}
    \label{tab:parameters}
    \begin{tabular}{|l|c|c|}
        \hline
        Parameter & TE & TM  \\
        \hline
        FWHM at Critical Coupling (GHz) & 2.5 & 7.0 \\
        FSR (GHz) & 438 & 426\\
        Propagation loss (dB/cm) & 2.5 & 7.1\\
        Intrinsic Q factor & \(1.5 \times 10^5\) & \(0.55 \times 10^5\) \\
        \(d \nu/dT\) at 785 nm (GHz/K) & 5.7 & 6.2 \\
        \(dn/dT\)  \((\times 10^{-5} \, \text{K}^{-1})\) & 2.9 & 3.26 \\
        \(n_{\text{AlGaN}}\) at 632 nm & 2.13 & 2.19 \\
        \(n_{\text{AlN}}\) at 632 nm & 2.05 & 2.10 \\
        \hline
    \end{tabular}
\end{table}

We mount the chips on a custom-made temperature-controlled stage to test the temperature tunability of the TE and TM modes resonance frequencies in the range of 20-30 $^\circ$C. Fig. \ref{fig:ring}(c) depicts the shift of a resonant peak of the TE mode (coupling gap=100nm) for different temperatures. In the investigated temperature range, the central frequency of the resonance changes linearly with temperature. The corresponding temperature coefficients are 5.7 and 6.2 GHz/K for TE and TM modes, respectively. These coefficients correspond to a thermo-optic coefficient of 2.9 $\times$ 10$^{-5}$ K$^{-1}$ and 3.6 $\times$ 10$^{-5}$ K$^{-1}$for  TE and TM polarizations in the \mbox{AlGaN} waveguide, which is comparable to the previously reported values for AlN (2.94$\times$ 10$^{-5}$ K$^{-1}$) and GaN (7.01 $\times$ 10$^{-5}$ K$^{-1}$) \cite{Watanabe2008} at 785 nm and 300K.

\begin{figure}[ht]
    \centering
    \includegraphics[width=0.45\textwidth]{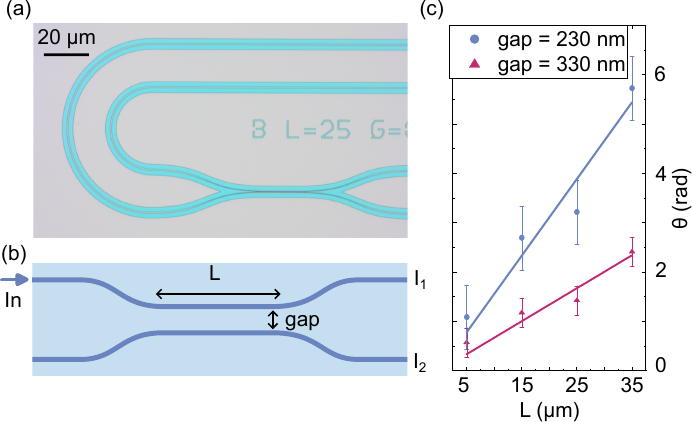}
    \caption{\textbf{Directional couplers.} (a) Optical microscope image of an AlGaN directional coupler. (b) Schematic illustration of the geometry, where the relevant parameters are presented. (c) Output intensity ratios, $\Theta$, across various coupling lengths and gap sizes, enabling the determination of the coupling constant for gaps measuring 230 and 330 nm.}
    \label{fig:dc}
\end{figure}

\subsection{Directional Couplers}

Directional couplers are essential components in photonic integrated circuits. They enable the transfer of electromagnetic energy between two or more waveguides that are in proximity. They can be combined to form optical switches or Mach-Zehnder interferometers, which are the fundamental constituents of most on-chip quantum technology applications  \cite{Zhou2022,Clements:16}. In our study, we develop \mbox{AlGaN} directional couplers linked to racetrack waveguides, as depicted in Fig. \ref{fig:dc}(a). We vary coupling gaps and lengths, as illustrated in Fig. \ref{fig:dc}(b). By injecting a 632 nm laser into one of the input ports, we measure the output intensity ratio: $\theta= \tan^{-1}(I_1/I_2)$. Our observations reveal a direct linear correlation between the coupling length and $\theta$, with linear fitting of the data providing the coupling constants. For coupling gaps of 230 nm and 330 nm, the coupling constants were determined to be 0.15 $\mu$m$^{-1}$ and 0.066 $\mu$m$^{-1}$, respectively. The observed uncertainties are primarily attributed to reflections at the input and output facets, which can be minimized by applying an anti-reflection coating. The well-functioning directional couplers show that our \mbox{AlGaN} platform is well-suited for building the required components of complex integrated photonic circuits. Further improvement is to be expected by optimizing the etching processes towards vertical sidewalls. 

\subsection{Phase Matching for Nonlinear Frequency Conversion and Pair Generation}
\begin{figure}[ht]
    \centering
    \includegraphics[width=\columnwidth]{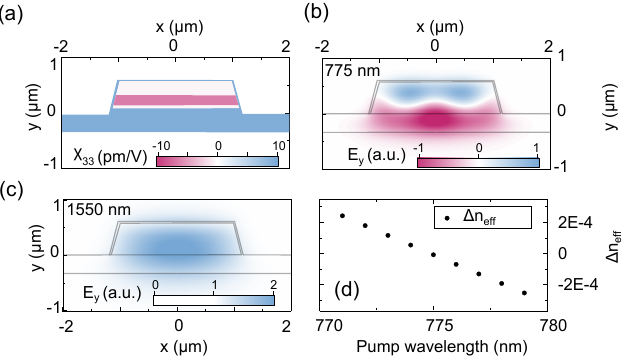}
    \caption{\textbf{Phase matching for SPDC.} An optimized heterostructure of Al$_x$Ga$_{1-x}$N for entangled photon pair generation at 1550 nm. (a) Second-order nonlinear susceptibility tensor element $\chi_{33}$, (b) vertical component of the electric field of the higher order TM mode at 775 nm, (c) the fundamental TM mode at 1550 nm, and  (d) effective index difference between these two modes as a function of pump laser wavelength. }
    \label{fig:phasematch}
\end{figure}
In nonlinear photonics applications, achieving a phase matching of the modes involving the nonlinear conversion process is critical for efficiency.  For integrated photonic waveguides, a common strategy is to match the effective indices of two distinct modes, typically a fundamental mode and a higher order mode \cite{AlGaNRing,Pernice2012}. These approaches, however, suffer from a weak overlap between the involved modes. The mode overlap integral for second harmonic generation and SPDC for degenerate photon pairs is defined as \cite{modeoverlap}
\begin{equation}
    \Gamma=\left|\int \chi^{(2)}: \mathbf{E_{\omega}}^2 \mathbf{E_{2\omega}} ^* \mathrm{~d} \Omega\right|,
\end{equation}
where $\chi^{(2)}$ is the second-order nonlinear susceptibility tensor, $\mathbf{E_{\omega}}$ and  $\mathbf{E_{2\omega}}$ are the normalized mode electric field amplitudes of the modes at $\omega$ and $2 \omega$, respectively. The integral is calculated over the mode volume $\Omega$. In the case of phase-matched TM modes for \mbox{AlGaN} waveguides, the dominating term in the overlap integral is
\begin{equation}
    \Gamma\approx \left|\int \chi_{33}E_{1y}^2 E_{2y} ^* \mathrm{~d} x\mathrm{~d} y\right|,
\end{equation}
where $\chi_{33}$ is the relevant component of $\chi^{(2)}$ for the interaction along the primary axis of the crystal in TM modes, and $E_{1y}$ and $E_{2y}$ are the y-components of the electric field amplitudes. Al$_x$Ga$_{1-x}$N alloys exhibit a sign change of the second-order nonlinear coefficient tensor component ($\chi_{33}$) at an alloy content of approximately $x$ = 0.65 \cite{AlGaN_suscept_meas}. We use this property to enhance the mode overlap.  We design a heterostructure to achieve phase matching between the fundamental TM mode at 1550 nm and a higher-order mode at 775 nm with identical polarization, as demonstrated in Fig. \ref{fig:phasematch}(a-c). This strategy facilitates efficient phase matching, enabled by the sign change of  $\chi_{33}$, which significantly increases the overlap integral between the pump and SPDC modes. Our concept also allows for fine-tuning phase matching by adjusting the pump wavelength. The resultant effective index difference,  $\Delta n_{eff}$, between the pump and down-conversion modes, which converge to zero at approximately 775 nm, is depicted in Fig. \ref{fig:phasematch}(d). For this simulation, we used a finite element solver (Comsol Multiphysics) to calculate the modes' effective indices and used a genetic algorithm (supplied by Matlab) to minimize the objective function, $(10^2\Delta n_{eff}-\Gamma)$. From bottom to top, optimized \mbox{AlGaN} layer thicknesses are 110/70/290/50 nm on 400 nm AlN on sapphire, while the alloy compositions are 0.55/0.42/0.65/0.66. The waveguide's top width is 1.98 $\mu$m, and the AlN layer is over-etched to a depth of 60 nm. Additionally, we define a 30 nm conformal AlN cladding. This particular layer design will likely exhibit some degree of strain relaxation, especially in the layers with low Al content. Changes in strain and piezoelectric polarization may impact the alignment of this calculation with experimental results. However, the overall viability of this method remains unaffected. The AlGaN/AlN heterostructure concept enables the engineering of compact SPDC sources and phase-matched ring resonators across a wide spectral range when coupled to a single-mode waveguide in a ring waveguide.

\section{Discussion}
We present a new material platform based on \mbox{AlGaN} heterostructures, which can be used to implement photonic devices with the core functionalities required to develop advanced photonic integrated chips. By epitaxially growing \mbox{AlGaN} on HTA-AlN templates on sapphire substrates, we alleviate the challenges related to crystal defects and dislocations at the sapphire-AlN interface and achieved an RMS surface roughness as low as 0.5 nm. The optical losses we quantify, 2.5 dB/cm for TE$_0$ mode and 7.1 dB/cm for TM$_0$ mode at 785 nm, outperform earlier measurements on aluminum nitride on sapphire waveguides in the visible spectrum. For instance, Lu et al. found losses of 5.3 dB/cm at 633 nm for the TE$_0$ mode \cite{AlNPlatformLu:18}. It is worth noting that we did not use a cladding material, which typically helps to reduce scattering loss further. By optimizing the layer sequence and growth recipes, as well as device fabrication, further reduction of losses will be possible.
Moreover, by measuring the thermo-optic coefficient and anisotropic refractive indices of the waveguides, we have gained insight into parameters relevant to applications such as nonlinear photon-pair generation and on-chip nonlinear conversion. The birefringent refractive index of our devices opens avenues for exploration, especially in its utilization for phase matching of TE and TM modes. Furthermore, the surface roughness of the epitaxial material is sufficiently low to integrate superconducting single-photon detectors \cite{SNSPD2}. Our work introduces \mbox{AlGaN} heterostructures as a novel platform for integrated photonics, offering a scalable and high-quality alternative to well-established materials.

\begin{acknowledgments}
We thank Torsten Petzke and Cornelia Neumann for assistance in AlGaN growth, Karina Ickert and Nico Sluka for their help in spin-coating and lithography, Joost Wartena, Adrian Runge, Mohammad Mohammadi, and Dominik Sudau for their assistance in thin-film depositions; Kevin Kunkel, Natalia Sabelfeld, Kai Gehrke, and Andreas Renkewitz for performing plasma etching processes; Alexander Külberg and Felix Eiche for laser cutting and marking; Uwe Spengler and Christine Münnich for cleaving, Olaf Krüger, Ralph-Stephan Unger and Ina Ostermay for fruitful discussions. We also thank Professor Arno Rauschenbetuel for providing access to his laboratory and the ring resonator measurement resources. This work received support from the European Research Council (ERC), Starting Grant project QUREP, No. 851810, the German Federal Ministry of Education and Research (BMBF, project DiNOQuant, No. 13N14921; project QPIS, No. 16KISQ032K; project QPIC-1, No. 13N15858), the Einstein Foundation Berlin (Einstein Research Unit: Perspectives of a quantum digital transformation: Near-term quantum computational devices and quantum processors), and Alexander von Humboldt Foundation in the framework of the Alexander von Humboldt Professorship endowed by the BMBF.
\end{acknowledgments}

\section*{contributions}S.G. led microfabrication, characterization, simulations, and manuscript writing; S.P. specialized in ring resonator characterization; T.P. conducted e-beam lithography and supervised the fabrication processes; T.K. and S.H. were responsible for \mbox{AlGaN} material growth, M.W. and T.S. led the project. All authors reviewed the manuscript.

\bibliography{apssamp}

\end{document}